\documentclass[10pt]{article}
\usepackage{graphicx}
\usepackage[cp1251]{inputenc}
\usepackage[russian]{babel}

\begin{document}
\twocolumn [ \noindent {\footnotesize\it  ISSN 1063-7737,
 Astronomy Letters, 2006, Vol. 32, No. 12, pp. 816--826.
 \copyright Pleiades Publishing, Inc., 2006.

\noindent Original Russian Text \copyright V.V. Bobylev, 2006,
published in Pis'ma v Astronomicheski$\check{\imath}$ Zhurnal,
2006, Vol. 32, No. 12, pp. 906--917.}

\vskip -4mm

\begin{tabular}{llllllllllllllllllllllllllllllllllllllllllllllll}
 & & & & & & & & & & & & & & & & & & & & & & & & & & & & & & & & & & & & & & & \\
\hline \hline
\end{tabular}

\vskip 1.2cm

 \centerline {\Large\bf Kinematics of the Gould Belt Based on Open Clusters}
 \bigskip
 \centerline {\large\bf V. V. Bobylev}
 \medskip
 \centerline {\it Pulkovo Astronomical Observatory, Russian Academy of Sciences,}
 \centerline {\it Pulkovskoe shosse 65, St.Petersburg, 196140 Russia}
 \centerline {\small Received April 13, 2006}

 \medskip
 \centerline {\small E-mail: {\tt vbobylev@gao.spb.ru}}

 \bigskip

{\noindent {\bf Abstract---}\small
 We have redetermined kinematic
parameters of the Gould Belt using currently available data on the
motion of nearby young ($\log t < 7.91$) open clusters, OB
associations, and moving stellar groups. Our modeling shows that
the residual velocities reach their maximum values of $-4$ km
s$^{-1}$ for rotation (in the direction of Galactic rotation) and
$+4$ km s$^{-1}$ for expansion at a distance from the kinematic
center of $\approx300$~pc. We have taken the following parameters
of the Gould Belt center: $R_\circ=150$ pc and
$l_\circ=128^\circ$. The whole structure is shown to move relative
to the local standard of rest at a velocity of $10.7\pm0.7$ km
s$^{-1}$ in the direction $l=274^\circ\pm4^\circ$ and
$b=-1^\circ\pm3^\circ$. Using the derived rotation velocity, we
have estimated the virial mass of the Gould Belt to be
$1.5\times10^6 M_\odot$.}
\bigskip


{\bf DOI}: 10.1134/S1063773706120036


\vskip 1cm

]

\section*{INTRODUCTION}

The Gould Belt is the nearest giant star--gas complex. It has been
firmly established that such complexes are regions of active star
formation. They are observed not only in our Galaxy (Efremov
1998), but also in other galaxies (Efremov 1989; Efremov and
Elmegreen 1998). The Gould Belt is a member of the older and more
massive local system of stars or the Local (Orion) Arm. This Belt
of bright stars was first identified by Herschel (1847), while the
Galactic coordinates of the pole of the great circle of the
celestial sphere along which the stars are grouped were determined
by Gould (1874). Based on a careful analysis of the spatial
distribution and density of OB stars from the Hipparcos catalogue
(1997), Torra et al. (2000) improved geometrical parameters of the
Gould Belt: the inclination to the Galactic plane, $16-22^\circ$,
and the direction of the line of nodes, $275-295^\circ$.

The Gould Belt also includes a system of nearby OB associations
(Lindblad et al. 1997; de Zeeuw et al. 1999). In the last
decade,more than a hundred T Tauri stars have been found in nearby
OB associations and open clusters owing to the spaceborne ROSAT,
Chandra, and XMM-Newton X-ray observations, the ground-based 2MASS
infrared photometric observations, and the astrometric Hipparcos
and Tycho-2 catalogues (Guillout et al. 1998; Mamajek et al. 2002;
Wichmann et al. 2003; Makarov 2003). These are late-type low-mass
pre-main-sequence stars with ages of several million years that
are identified by a number of characteristic signatures . lithium
overabundance, X-ray emission, etc. These stars are of interest to
us in that they expand the list of stars for studying the
kinematics of the Gould Belt.

HI clouds (Olano 1982; P\"{o}ppel and Marronetti 2000) and H2
molecular cloud complexes (Perrot and Grenier 2003) are also
associated with the Gould Belt.

The structure of interstellar tenuous hot gas in the immediate
solar neighborhood with a radius of $200-300$ pc is also closely
associated with the Gould Belt. The so-called Local Bubble was
found here (Sfeir et al. 1999). In the opinion of Bergh\"{o}fer
and Breitschwerdt (2002), the most realistic theory for the origin
of the Local Bubble is the hypothesis about the explosions
$\sim20$ supernovae in the past $10-20$ Myr. Seven neutron stars
that are the remnants of these supernovae have already been
discovered in the solar neighborhood (Popov et al. 2003).

Several formation scenarios for the Gould Belt have been suggested
to date. According to the first scenario, the collision of
high-velocity HI clouds with the Galactic disk led to its
formation (Franko et al. 1988; Comer\'{o}n and Torra 1992, 1994).
According to the second scenario, a supernova explosion led to the
formation of the Gould Belt (Olano 1982; P\"{o}ppel and Marronetti
2000). According to the third scenario (Olano 2001), the formation
of the Gould Belt is a stage in the kinematic evolution of the
local system of stars.

Olano (1982) considered a gasdynamic model for the formation of
the Gould Belt with a significant initial expansion velocity of
the primordial cloud. However, because of the braking of neutral
hydrogen caused by the drag of the ambient medium, the expansion
velocity dropped to zero. According to this model, it is the
zero-velocity boundary that delineates the outer boundary of the
Gould Belt (an ellipse with semi-axes of 364 pc and 211 pc was
found). The age of the Gould Belt was found using this model to be
$30\times10^6$ yr. The total mass of the expanding hydrogen was
estimated to be $1.2\times10^6 M_\odot$; the parameters of the
system's center were determined: $l_\circ = 131^\circ$ and
$R_\circ = 166$ pc.

Lindblad (2000) suggested a model for the proper rotation of the
Gould Belt with an angular velocity,
 $\omega_\circ = -24$ km s$^{-1}$ kpc$^{-1}$, that coincides in direction with the Galactic
rotation,
 $\omega'_\circ = 77$ km s$^{-1}$ kpc$^{-2}$, as well as
the system's expansion with
 $\rho_\circ = 20$ km s$^{-1}$ kpc$^{-1}$ and
 $\rho'_\circ = -117$ km s$^{-1}$ kpc$^{-2}$ for the
inferred direction of the rotation/expansion center $l_\circ =
127^\circ$ and the adopted $R_\circ = 166$ pc. This model takes
into account the disk inclination to the Galactic plane,
$20^\circ$ The model was constructed based on the results of
Hipparcos data analysis, which are reflected in the papers by
Comer\'{o}n (1999) and Torra et al. (2000). In contrast to the
model by Olano (1982), the model by Lindblad (2000) explains the
flat shape of the Gould Belt by the presence of a significant
angular momentum.

By modeling the dynamical evolution based on HI and H2 clouds,
Perrot and Grenier (2003) showed that the current main kinematic
parameters of the Gould Belt are virtually independent of the
direction of initial rotation. Previously (Bobylev 2004b), we
developed Lindblad's approach for the case where the distances are
calculated accurately. Using the residual space velocities of
individual stars belonging to nearby OB associations, we
determined the following kinematic parameters of the Gould Belt:
the proper rotation in the direction of Galactic rotation,
 $\omega_\circ =-23.1\pm2.2$ km s$^{-1}$ kpc$^{-1}$ and
 $\omega'_\circ= 31.3\pm6.5$ km s$^{-1}$ kpc$^{-2}$, as well as
the expansion with
 $k_\circ= 14.0\pm2.2$ km s$^{-1}$ kpc$^{-1}$
and
 $k_\circ= -27.3\pm6.5$ km s$^{-1}$ kpc$^{-2}$ for the
inferred coordinates of the center
 $l_\circ= 128^\circ$ and
 $R_\circ= 150$ pc.

The Gould Belt is the youngest constituent of the local system of
stars. Barkhatova et al. (1989) analyzed the properties of several
nearby single open clusters of various ages and some of the open
cluster complexes and hypothesized that they belong to a
higher-order system, the Supercomplex. The Supercomplex was shown
to have a residual rotation with an angular velocity
 $\omega_\circ\approx-12$ km s$^{-1}$ kpc$^{-1}$.

According to the theory by Olano (2001), the gas out of which the
local system of stars was formed had a high initial velocity
($\approx50$ km s$^{-1}$). The collision of the gas cloud with a
spiral density wave led to its fragmentation. In this model, such
clusters as the Hyades, Pleiades, Coma, and Sirius are considered
as fragments of once a single complex and contraction of the
central regions of the parent cloud led to the formation of the
Gould Belt.

The Gould Belt is one of the key objects for understanding the
evolution of the local system of stars, since its constituents
(gas, stars, and clusters) have a low velocity dispersion and have
not moved far away from their birthplace. Therefore, improving the
kinematic parameters of the Gould Belt is topical.

At present, an unprecedented (in completeness) catalog of 652 open
clusters presented by Kharchenko et al. (20065a) and analyzed by
Piskunov et al. (2006) has appeared. The goal of this paper is to
redetermine kinematic parameters of the Gould Belt based on
currently available data on individual clusters. The advantage of
this approach over the method with individual stars is that the
age and distance estimates for clusters are more reliable.

\section*{DATA}

We will designate the catalog of 652 open clusters as COCD
(Kharchenko 2001; Kharchenko et al. 2005a, 2005b; Piskunov et al.
2006). The young open cluster complex in the solar neighborhood
with an age of $\log t < 7.9$ designated by Piskunov et al. (2006)
as OCC1 provides a basis for our work list.

In COCD, the random errors of the mean cluster proper motions are
small, since the ASCC-2.5 catalog (Kharchenko 2001) compiled from
Hipparcos, Tycho-2 (Hog et al. 2000), and several other sources
was used to determine them.Data from the catalog by
Barbier-Brossat and Figon (2000) served as a basis for deriving
the mean radial velocities of the COCD clusters. At present, more
accurate stellar radial velocities, the OSACA catalog (Bobylev et
al. 2006; Gontcharov 2006), are available; their main peculiarity
is that they have all been reduced to a single standard. The OSACA
catalog combines several major catalogs of stellar radial
velocities:WEB (Duflot et al. 1995), Barbier-Brossat and Figon
(2000), Nordstr\"{o}m et al. (2004), Famaey et al. (2005), as well
as several fragmentary observations taken from $\approx500$
publications.

Young open clusters, which are compact, gravitationally bound
systems, are currently believed to be members of larger-scale, but
gravitationally less bound structures, associations. The list of
OB associations belonging to the Gould Belt structure and the list
of identified Hipparcos stars (probable members of these
associations) are presented in de Zeeuw et al. (1999).We compared
the two lists of Hipparcos stars, more specifically, the list of
OCC1 cluster stars and the list of de Zeeuw et al. (1999). This
comparison showed that the Orion complex is fully represented in
the OCC1 list of open clusters, while the Sco-Cen complex is
barely reflected in this list. To compile the most complete list
of stellar groupings belonging to the Gould Belt structure, we
made several additions. These are described in Section 3.1 and are
related mainly to the Sco-Cen complex.

In comparison with our previous paper (Bobylev 2004b), in this
paper we made several significant additions for the stars of the
Sco-Cen complex using the list of de Zeeuw et al. (1999) and the
OSACA database. In our analysis, we included the open cluster
Chamaeleontis, the data for which were taken from Sartori et al.
(2003), the Hipparcos catalog, and the OSACA database. We used the
latest data for 25 TWA cluster stars from Mamajek (2005).

\section*{MODELS}

In this paper, we use a rectangular Galactic coordinate system
with the axes directed away from the observer toward the Galactic
center ($l=0^\circ$, $b=0^\circ$, the $x$ axis), along the
Galactic rotation ($l=90^\circ$, $b=0^\circ$, the $y$ axis), and
toward the North Galactic Pole ($b=90^\circ$, the $z$ axis).

We apply the equations derived from Bottlinger's standard formulas
(Ogorodnikov 1965) by assuming the existence of a common kinematic
center for rotation and expansion using two terms of the Taylor
expansion of the angular velocity for rotation and the analogous
parameter for expansion (Lindblad 2000; Bobylev 2004b). The
equations are
$$
\displaylines{\hfill
  V_r= -u_{\odot}\cos b\cos l-\hfill\llap{(1)}
\cr\hfill
      -v_{\odot}\cos b\sin l-w_{\odot}\sin b+\cos^2 b k_\circ r+
\hfill\cr\hfill
  +(R-R_\circ)(r\cos b-R_\circ\cos (l-l_\circ))\cos b  k'_\circ-
\hfill\cr\hfill
 -R_\circ (R-R_\circ)\sin (l-l_\circ) \cos b
 \omega'_\circ,\hfill\cr
\hfill 4.74 r \mu_l\cos b=u_{\odot}\sin l-v_{\odot}\cos l -
\hfill\llap(2)\cr\hfill
  -(R-R_\circ)(R_\circ\cos (l-l_\circ)-r\cos b)\omega'_\circ+
\hfill\cr\hfill
 +r\cos b \omega_\circ+R_\circ(R-R_\circ)\sin (l-l_\circ)) k'_\circ,\hfill
 \cr
\hfill 4.74 r \mu_b=u_{\odot}\cos l\sin b+ \hfill\llap(3)
\cr\hfill
 +v_{\odot}\sin l \sin b-w_{\odot}\cos b-\cos b\sin b k_\circ r-
\hfill\cr\hfill
  -(R-R_\circ)(r\cos b-R_\circ\cos (l-l_\circ))\sin b k'_\circ+
\hfill\cr\hfill
 +R_\circ(R-R_\circ)\sin (l-l_\circ)\sin b\omega'_\circ.\hfill
 }
$$
Here, the coefficient 4.74 is the quotient of the number of
kilometers in an astronomical unit by the number of seconds in a
tropical year, $r = 1/\pi$ is the heliocentric distance of the
star, $R_\circ$ is the distance from the Sun to the kinematic
center, $R$ is the distance from the star to the center of
rotation, $l_\circ$ is the direction of the kinematic center, and
$u_\odot$, $v_\odot$, and $w_\odot$ are the components of the
peculiar solar motion. The stellar proper-motion components $\mu_l
\cos b$ and $\mu_b$ are in mas yr$^{-1}$, the radial velocity
$V_r$ is in km s$^{-1}$, the parallax $\pi$ is in mas, and the
distances $R$, $R_\circ$, and $r$ are in kpc. The quantity
$\omega_\circ$ is the angular velocity and $k_\circ$ is the radial
expansion/contraction velocity of the stellar system at distance
$R_\circ$; the parameters $\omega'_\circ$ and $k'_\circ$ are the
corresponding derivatives. The distance $R$ can be calculated
using the expression
$$
\displaylines{\hfill
 R^2=(r\cos b)^2
 -2R_\circ r\cos b\cos (l-l_\circ)+R^2_\circ.\hfill
 }
$$
The system of conditional equations (1)--(3) contains seven
unknowns: $u_\odot$, $v_\odot$, $w_\odot$, $\omega_\circ$,
$\omega'_\circ$, $k_\circ$, and $k_\circ$, to be determined by the
least-squares method. The lefthand sides of Eqs. (1)--(3) are
assumed to be freed from the differential Galactic rotation.

We also apply a more general method based on the analysis of the
linear rather than angular residual velocities, which we
approximate by a series of the form
$$
\displaylines{\hfill
  V_i= V_{\circ, i}+a_i (R-R_\circ)+ b_i (R-R_\circ)^2, \hfill\llap(4)
 }
$$
where $i = R, \theta$ and the residual expansion, $V_R$, and
rotation, $V_\theta$, velocities must be obtained by decomposing
the observed velocities (components $U$ and $V$) into the radial
and tangential components at fixed parameters of the center,
$l_\circ$ and $R_\circ$. The system of conditional equations (4)
contains three sought-for unknowns: $V_{\circ,i}$, $a_i$, and
$b_i$, to be determined by the least-squares method.

\section*{RESULTS}
\subsubsection*{Improving the Work List of Open Clusters}

Using the OSACA catalog of radial velocities, we improved the mean
radial velocities for the clusters ASCC 16 and ASCC 18. Two stars,
HIP25288 and HIP25302, were used for ASCC 16 in COCD. We suggest
adding two more stars, HIP25163 and HIP25235, for which radial
velocities are available in OSACA. We calculated the new mean
radial velocity of the cluster using three stars (HIP25288 was
excluded). The new mean radial velocity for ASCC 16 affected
significantly the space velocities $U$ and $V$, since the
difference between the $V_r$ determinations was $\sim15$ km
s$^{-1}$.

The radial velocity of ASCC 18 in COCD was determined using the
star HIP25378. We also suggest adding the star HIP25567.

We suggest attributing the open cluster ASCC114 to the Gould Belt
structure. Unfortunately, only one star was used to calculate the
mean radial velocity.

Since the age of the open cluster NGC 2516 is $\log t = 7.91$, it
was not included in OCC1 by Piskunov et al. (2006). However, based
on the results by Torra et al. (2000) and Bobylev (2004b), we
attribute it to the Gould Belt complex.

Piskunov et al. (2006) estimated the probability ($P_t$) that the
clusters Stock 23 and Melotte 20 belong to the Gould Belt
structure as 25\% and 2\%, respectively. We do not include them in
the main list to determine the kinematic parameters of the Gould
Belt; their membership in the Gould Belt structure is discussed
below. The open cluster ASCC 89 has $P_t=42$\%. However, since
this cluster is located fairly far from the presumed kinematic
center, in the fourth quadrant ($x=415$ pc, $y=-250$ pc, it is
outside the boundary in Fig.~2), and has a high residual velocity
$W$, we do not consider it either.

Tables 1 and 2 give the space velocities $U,V,W$ calculated for
our 49 clusters that were corrected for the differential Galactic
rotation. To calculate the velocities $U,V,W$, which are listed in
Table 2, we took all of the input data, more specifically, the
coordinates of the centers $\alpha, \delta$, the velocity
components $\mu_\alpha \cos \delta$, $\mu_\delta$, $V_r$, their
random errors $e_{\mu_{\alpha}\cos\delta}$, $e_\mu\delta$,
$e_{V_r}$, and the mean distances $d$ from COCD without any
changes.

We estimated the errors in the velocities $U,V,W$, which are given
in Tables 1 and 2, by assuming that $e_\pi/\pi = 0.1$. The list of
stars for Table 1 was compiled under the condition $\pi>1.5$ mas.
The asterisks mark the four clusters for which the random errors
in one of the velocity components $U,V,W$ exceed 5 km s$^{-1}$.
Large errors are almost always, except the Sco-Cen complex, caused
by large random errors in the radial velocities. For this reason,
we did not use the parameters of the cluster Tr 10 from COCD, for
which the formal error $e_{V_r}$ reaches 15 km s$^{-1}$. Instead
we determined the mean input parameters using the list of stars
for Tr 10 that we compiled based on the papers by de Zeeuw et al.
(1999), Robichon et al. (1999), and the OSACA catalog.

At present, 13 stars are attributed to the moving cluster $\beta$
Pic (Ortega et al. 2004). We determined the parameters of the
cluster listed in Table 1 using the list of six candidates from
Song et al. (2003). These stars form a compact group in the second
Galactic quadrant (see Fig. 2).

In our analysis, we included the set of very young and nearby
(within 50 pc of the Sun) stars from Makarov (2003, designated as
set $XY$) and Wichmann et al. (2003, designated as set ZAMS) and
OSACA radial velocities. These are late-type low-mass
pre-main-sequence stars with ages of several million years
identified by lithium overabundance and X-ray emission (based on
the ROSAT survey).

Depending on the distance, we divided the four associations US,
UCL, LCC, and Lac OB1, which contain a fairly large number of
stars, in two. The main goal of this formal division is to
increase the amount of data for solving Eqs. (1)--(3) and (4). The
following distances were used as the boundary ones: 143, 140, 110,
and 300 pc for US, UCL, LCC, and Lac OB1, respectively. For the
Lac OB1 association, the difference between the mean distances of
parts a and b turned out to be largest, $\sim$220 pc (see Table
1).

\subsubsection*{The Motion Relative to the Local Standard of Rest}

Based on the linear Ogorodnikov.Milne model, we determined the
mean peculiar solar motion using 49 clusters at a mean
heliocentric distance of 312 pc, $V_\odot=19.69\pm0.71$ km
s$^{-1}$ in the direction $L_\odot=60^\circ\pm2^\circ$ and
$B_\odot=21^\circ\pm2^\circ$. The components of this motion were
found to be ($u_\odot,v_\odot,w_\odot)=(9.26, 15.93, 6.96)\pm
(0.72,0.72,0.60)$ km s$^{-1}$.

Using these clusters, we determined the mean motion of the Gould
Belt complex relative to the local standard of rest,
$V_G=10.7\pm0.7$ km s$^{-1}$, in the direction
$L_G=274^\circ\pm4^\circ$ and $B_G=-1^\circ\pm3^\circ$. As the
standard solar motion relative to the local standard of rest, we
took (10.0, 5.3, 7.2) km s$^{-1}$ from Dehnen and Binney (1998).

\subsubsection*{Correction for the Galactic Rotation}

Figure 1 shows the dependence of the velocity U of Gould Belt
stars on coordinate $y$: $U=U_\circ+Y(dU/dY)$, where
 $U_\circ=-9.3\pm0.7$ km s$^{-1}$. The slope of the line $dU/dY=26.6$ km s$^{-1}$
 kpc$^{-1}$ corresponds to our correction $-(B-A)=-\omega_\circ$.
We use $A=13.7\pm0.6$ km s$^{-1}$ kpc$^{-1}$ and $B=-12.9\pm0.4$
km s$^{-1}$ kpc$^{-1}$ found previously (Bobylev 2004a). In the
above paper, we determined the following angular velocity
parameters for the Galactic rotation: $\omega_\circ=-28.0\pm0.6$
km s$^{-1}$ kpc$^{-1}$, $\omega'_\circ=+4.17\pm0.14$ km s$^{-1}$
kpc$^{-2}$, $\omega''_\circ=-0.81\pm0.12$ km s$^{-1}$ kpc$^{-3}$
at $R_\circ=7.1$ kpc. Comparison of the two methods of correction
for the differential Galactic rotation shows that there are no
significant differences between the two methods in the solar
neighborhood $\approx500$ pc in radius under consideration. We
then applied a correction for the Galactic rotation by the simpler
first method, using the Oort constants $A$ and $B$.

As can be seen from Fig. 1, there is a local feature in the range
$-200<Y<0$ pc that is related to the proper rotation of the Gould
Belt stars. An examination of Fig. 1 shows that the corrections
for the Galactic rotation should undoubtedly be applied in the
range of distances under consideration.

\subsubsection*{Residual Rotation and Expansion of the Gould Belt}

Figures 2--5 present the residual cluster velocities. The
velocities shown in Fig. 4 were freed from the differential
Galactic rotation. The velocities shown in Figs. 2, 3, and 5 were
freed from both the differential Galactic rotation and the linear
motion $(u_\odot,v_\odot,w_\odot)=(9.23,15.93,6.96)$ km s$^{-1}$
found above.

Figure 2 shows the residual velocity vectors $U$ and $V$ in
projection onto the Galactic $xy$ plane and the residual velocity
vectors $U$ and $W$ in projection onto the Galactic $xz$ plane.
The circles mark the clusters that we did not use to determine the
rotation/expansion parameters for various reasons. The reasons are
the following: only one star was used to determine the radial
velocity of ASCC 114; there are common stars for calculating the
mean radial velocity for IC 348 and Per OB2, the situation is
similar for Col 132 and Col 121; and the open cluster Melotte 20
has a low probability of its membership in the Gould Belt
structure.

Figure 3 presents the residual velocities $U'$ as a function of
the coordinate $x'$ calculated at $l_\circ=160^\circ$. This figure
shows the dependence $dU'/dX'=17.9$ km s$^{-1}$ kpc$^{-1}$ that
corresponds to the first main root of the deformation tensor that
was estimated using the linear Ogorodnikov-Milne model from the
residual cluster velocities. We see from the figure that, apart
from the linear trend, there is a wave with a period of $\sim300$
pc. We interpret the presence of this wave as the existence of a
derivative (in linear velocities) that is related mainly to the
expansion effect.

Figure 4 shows the two-dimensional residual velocity $UVW$
distributions for the open cluster complex belonging to the Gould
Belt.

Figure 5 presents the residual expansion, $V_R$, and rotation,
$V_\theta$, velocities as a function of the distance from the
kinematic center $R$; the decomposition was performed for
 $l_\circ=128^\circ$ and $R_\circ=150$ pc. The parameters that we determined by
solving the system of equations (1)--(3) are most reliable:
$(u_\odot,v_\odot,w_\odot)=(9.6,17.3,7.0)\pm(0.9,0.9,0.6)$ km
s$^{-1}$ and
$$
\displaylines{\hfill
  \omega_\circ=-18.8\pm4.9~\hbox{km s$^{-1}$ kpc$^{-1}$},     \hfill\llap(5)\cr\hfill
 \omega'_\circ=+32.9\pm16.4~\hbox{km s$^{-1}$ kpc$^{-2}$},\hfill\cr\hfill
       k_\circ=+22.1\pm4.9~\hbox{km s$^{-1}$ kpc$^{-1}$},     \hfill\cr\hfill
      k'_\circ=-58.3\pm16.4~\hbox{km s$^{-1}$ kpc$^{-2}$},\hfill
}
$$
at $l_\circ=128^\circ$ and $R_\circ=150$ pc. We used parameters
(5) to construct the rotation and expansion curves in Fig. 5
(curves 1). For each curve 1 in Fig. 5, the thin lines indicate
 $1\sigma$ confidence regions.

Figure 5a also presents the expansion curves 2 and 3 whose
parameters were found from the linear velocities using Eq. (4). We
used all 49 points to determine the parameters of curve 2:
$$
\displaylines{\hfill
  V_{\circ R}= 4.1\pm2.1~\hbox{km s$^{-1}$},     \hfill\cr\hfill
          a_R=-19\pm22~\hbox{km s$^{-1}$ kpc$^{-1}$},     \hfill\cr\hfill
          b_R=  8\pm49~\hbox{km s$^{-1}$ kpc$^{-1}$}.   \hfill
}
$$
The parameters of curve 3 were determined for the constraint
$R<375$ pc (27 points):
$$
\displaylines{\hfill
  V_{\circ R}= 4.3\pm2.1~\hbox{km s$^{-1}$},     \hfill\cr\hfill
          a_R= 49\pm40~\hbox{km s$^{-1}$ kpc$^{-1}$},     \hfill\cr\hfill
          b_R=-40\pm20~\hbox{km s$^{-1}$ kpc$^{-1}$}.   \hfill
}
$$
In contrast to curves 1 and 3, curve 2 describes the process where
the expansion decelerates immediately. All of the expansion curves
found show that the expansion is more local than the rotation.

The parameters of the rotation curve 2 in Fig. 5b were found from
the linear velocities using Eq. (4):
$$
\displaylines{\hfill
  V_{\circ \theta}=-1.5\pm1.9~\hbox{km s$^{-1}$},     \hfill\cr\hfill
          a_\theta=-23\pm20~\hbox{km s$^{-1}$ kpc$^{-1}$},     \hfill\cr\hfill
          b_\theta=+40\pm45~\hbox{km s$^{-1}$ kpc$^{-1}$}.   \hfill
}
$$
As can be seen from Fig. 5b, there is good agreement between the
residual rotation curves for the Gould Belt obtained by the two
different methods.

Analysis of the derived curves indicates that the residual
velocities are determined reliably only in a close neighborhood,
$R\approx R_\circ$, and are equal to $-2.8\pm0.7$ km s$^{-1}$ for
rotation and $3.3\pm0.7$ km s$^{-1}$ for expansion. The maximum
values are reached by these velocities at slightly different
distances ($R_{max}$) from the kinematic center and are equal to
$-4.3\pm1.9$ km s$^{-1}$, $R_{max}=360$ pc for rotation and
$4.1\pm1.4$ km s$^{-1}$, $R_{max}=265$ pc for expansion. Based on
the maximum rotation velocity found, we obtained a virial mass
estimate for the Gould Belt, $1.5\times10^6 M_\odot$.

We also expanded the list of individual stars that belong to the
clusters and associations under consideration. This list contains
a total of $\sim$700 Hipparcos star, more that 400 stars of which
are members of the Sco-Cen complex. The residual velocities of 658
stars satisfy the criterion $\sqrt{U^2+V^2+W^2}<60$ km s$^{-1}$.
Based on these 658 individual stars, we found the following
kinematic parameters of the Gould Belt by solving the system of
equations (1)--(3):
$(u_\odot,v_\odot,w_\odot)=(6.02,16.71,5.62)\pm(0.23,0.23,0.20)$~km
s$^{-1}$ and
$$
\displaylines{\hfill
  \omega_\circ=-21.4 \pm1.6~\hbox{km s$^{-1}$ kpc$^{-1}$}, \hfill\llap(6)\cr\hfill
 \omega'_\circ=+25.7 \pm5.0~\hbox{km s$^{-1}$ kpc$^{-2}$}, \hfill\cr\hfill
       k_\circ=+25.6 \pm1.6~\hbox{km s$^{-1}$ kpc$^{-1}$}, \hfill\cr\hfill
      k'_\circ=-50.5 \pm5.0~\hbox{km s$^{-1}$ kpc$^{-2}$}. \hfill
}
$$
The question of whether the expansion of the Gould Belt stars can
be related solely to the peculiarities of determining the stellar
radial velocities has long been discussed. To test this
assumption, we solved the system of only two equations, (2) and
(3), (without involving the stellar radial velocities) using the
same set of individual stars. The results are the following:
$(u_\odot,v_\odot,w_\odot)=(8.42,15.40,6.31)\pm(0.27,0.29,0.15)$~km
s$^{-1}$ and
$$
\displaylines{\hfill
  \omega_\circ=-20.9 \pm 1.2~\hbox{km s$^{-1}$ kpc$^{-1}$}, \hfill\llap(7)\cr\hfill
 \omega'_\circ=+33.8 \pm 4.2~\hbox{km s$^{-1}$ kpc$^{-2}$}, \hfill\cr\hfill
       k_\circ=+28.7 \pm 3.6~\hbox{km s$^{-1}$ kpc$^{-1}$}, \hfill\cr\hfill
      k'_\circ=-50.4 \pm12.0~\hbox{km s$^{-1}$ kpc$^{-2}$}. \hfill
}
$$
Comparison of parameters (7) found with parameters (6) shows that
the expansion is actually a kinematic effect.

\subsubsection*{The Residual Velocity Ellipsoid}

Table 3 gives the principal semi-axes of the residual velocity
ellipsoid $\sigma_{1,2,3}$ and their directions $l_{1,2,3}$ and
$b_{1,2,3}$. In the first row of Table. 3, the parameters were
determined from the velocities that were freed from the peculiar
solar motion and the differential Galactic rotation. The axial
ratios are 8:4:3.

The second row of this table gives the parameters determined from
the cluster velocities that were also freed from the rotation
velocity and contraction of the Gould Belt found above (parameters
(5)). In this case, the axial ratios are 8:6:4, i.e., the residual
velocity distribution is closer to a spherical one.

There is an appreciable decrease in the residual velocity
$\sigma_1$. On the whole, we conclude that allowance for the
rotation velocities and contraction of the Gould Belt that we
found brings the residual velocities $\sigma_{1,2,3}$ closer to
the ``cosmic dispersion''. Since we solved the plane problem
(without including the inclination of the Gould Belt to the $z$
axis), the relationship of the residual velocities found (the
second row of the table) to the Gould Belt remains noticeable:
$b_1=-15^\circ$ at $l_1=188^\circ$; consequently, the true cosmic
dispersion must be even lower.

\subsubsection*{Probable Members of the Gould Belt Structure}

Figure 6 shows the probability density distribution of the
residual $UV$ velocities. The residual velocities shown in Fig. 4
serve as the input data for this distribution. We applied an
adaptive Gaussian smoothing method where the Gaussian width at
each point of the map depends on the density of the data
distribution near this point. A detailed description of the
algorithm for determining the probability density can be found in
Skuljan et al. (1999) and Bobylev et al. (2006). The contour lines
in Fig. 6 are given at 10\% steps. The positions of two open
clusters, Melotte 20 ($\alpha$ Per) and Stock 23, are shown. For
all the remaining clusters listed in Tables 1 and 2, the
probability exceeds 20\%. In particular, this also applies to such
clusters as $\sigma$ Ori, Platais 6, Col 135, Col 140, NGC 2451b,
and IC 2602, for which Piskunov et al. (2006) obtained low
probabilities $P_t$ (5\%, 1\%, 11\%, 15\%, 19\%, and 13\%,
respectively).

The low probability that we found for Stock 23 may be related to
an uncertainty in the radial velocity, since it was calculated
using only two stars.

The open cluster Melotte 20 is located at a distance of only 71 pc
from the presumed kinematic center. Piskunov et al. (2006)
estimated its age to be 35 Myr. The mean radial velocity was
calculated using 66 stars. Therefore, this cluster is of
considerable interest in studying the kinematics of the Gould Belt
and its evolution. However, including it in the data set for
determining the rotation and expansion parameters leads to an
appreciable increase in the error of a unit weight.

\section*{DISCUSSION}

The residual rotation velocity of the Gould Belt that we found,
$V_\theta=-2.8\pm0.7$ km s$^{-1}$ (at $R=150$ pc), is in complete
agreement with the results of the analysis (based on an
independent method) of the motion of OCC1, the youngest open
cluster complex, by Piskunov et al. (2006), who found the mean
tangential velocity to be ${\overline T}_l=-2.9\pm0.7$ km
s$^{-1}$. The discrepancy between the two approaches in estimating
the angular velocity $\omega_G$ is attributable to the use of
different parameters for the kinematic center.

The difference between our residual expansion velocity of the
Gould Belt $V_R=3.3\pm0.7$ km s$^{-1}$ (at $R=150$ pc) and the
velocity $V_r=2.1\pm1.1$ km s$^{-1}$ for OCC1 undoubtedly stems
from the fact that Piskunov et al. (2006) did not consider the
Sco.Cen complex (US, UCL, LCC), which has the dominant influence
on the determination of the expansion effect.

Our estimate of the cosmic dispersion for Gould Belt stars, (5.29,
4.28, 2.75) km s$^{-1}$, with respect to the first axis is smaller
than the estimate by Piskunov et al. (2006), (7.2, 4.1, 2.6) km
s$^{-1}$, obtained for the young open cluster complex with age
$\log t = 6.7$. This reduction in the cosmic dispersion can be
explained by allowance for the rotation and expansion parameters
of the Gould Belt that we found.

Comparison of the parameters derived from mean cluster velocities
(solution (5)) and from individual stars (solution (6)) shows
that, in the former case, the maximum velocities are found to be a
factor of $\approx 1.5$ lower than those in the latter case. This
difference is probably related to a difference between the
distance scales.We believe the photometric distance estimates
obtained by analyzing open clusters (particularly for distant
clusters) to be more reliable; therefore, we prefer solution (5).

The kinematic parameters (5) that we found are generally in
agreement with the results of Lindblad (2000), Perrot and Grenier
(2003), and Bobylev (2004b).

\section*{CONCLUSIONS}

Thus, we have performed a purposeful work on compiling the most
complete list of nearby young open clusters, OB associations, and
moving stellar groups that belong to the Gould Belt structure. In
particular, we showed that the probabilities that the clusters
Melotte 20 and Stock 23 belong to the Gould Belt structure are
$\sim$20\% and 10\%, respectively; the recently discovered
(Kharchenko et al. 2005b) open cluster ASCC 114 has a probability
of more than 20\%. The stellar radial velocity data for such
clusters as Platais 6, Alessi 5, NGC 2451b, ASCC 18, ASCC 114, and
Stock 23 must be improved.We used the OSACA database to improve
the radial velocities for a fairly large number of nearby stars.
In particular, we improved the mean radial velocities of the open
clusters ASCC 16 and ASCC 18.

The errors in the stellar radial velocities were shown to make the
dominant contribution to the errors in the space velocities
$U,V,W$. The nearby clusters of the Sco-Cen complex (US, UCL, LCC)
with a large number of members, where the random errors in the
stellar proper motions have the dominant influence, constitute an
exception. Thus, the task of improving the stellar radial
velocities remains topical; the prospect for its breakthrough
accomplishment is related to the accomplishment of the future GAIA
space mission.

To determine the kinematic parameters of the Gould Belt, we used
data on 45 clusters and OB associations. We took the data for 31
open clusters directly from COCD (Piskunov et al. 2006) and
obtained the data for 14 added clusters, moving groups, and
associations based on various sources. The mean age of the sample
is $\approx$32 Myr; the clusters are located within $\approx$500
pc of the Sun. The entire Gould Belt structure was shown to be
involved in several motions. First, apart from the involvement in
the overall rotation of the Galaxy, the complex as a whole moves
relative to the local standard of rest at a velocity of
$10.7\pm0.7$ km s$^{-1}$ in the direction $l=274^\circ\pm4^\circ$
and $b=-1^\circ\pm3^\circ$. Second, there is residual rotation and
expansion of the system. As the parameters of the kinematic
center, we took $l_\circ=128^\circ$ and $R_\circ=150$ pc that we
determined previously (Bobylev 2004b). Our modeling showed that
the residual velocities are reliably determined in a close
neighborhood, $R\approx R_\circ$, and are equal to $-2.8\pm0.7$ km
s$^{-1}$ for rotation (in the direction of Galactic rotation) and
$+3.3\pm0.7$ km s$^{-1}$ for expansion. The maximum values are
reached by these velocities at a distance from the kinematic
center of $\approx$300 pc and are $-4.3\pm1.9$ km s$^{-1}$ for
rotation and $+4.1\pm1.4$ km s$^{-1}$ for expansion. Based on the
rotation velocity found, we obtained a virial mass estimate for
the Gould Belt, $1.5\pm10^6 M_\odot$.

\subsection*{ACKNOWLEDGMENTS}

I am grateful to the referees for helpful remarks that contributed
to an improvement of the paper. I wish to thank G.~A.~Gontcharov,
who timely provided the last version of the OSACA catalog of
radial velocities, and A.~T.~Bajkova, who constructed the
probability density distribution of residual velocities. This work
was supported by the Russian Foundation for Basic Research
(project no. 05--02--17047).

\subsection*{REFERENCES}

 {
 1. M. Barbier-Brossat and P. Figon, Astron. Astrophys., Suppl.
Ser. {\bf 142}, 217 (2000).

2. K. A. Barkhatova, L. P. Osipkov, and S. A. Kutuzov, Astron. Zh.
{\bf 66}, 1154 (1989) [Sov. Astron. 33, 596 (1989)].

3. T.W. Bergh\"{o}fer and D. Breitschwerdt, Astron. Astrophys.
{\bf 390}, 299 (2002).

4. V. V. Bobylev, Pis'ma Astron. Zh. {\bf 30}, 185 (2004a)
[Astron. Lett. {\bf 30}, 159 (2004a)].

5. V. V. Bobylev, Pis'ma Astron. Zh. {\bf 30}, 861 (2004b)
[Astron. Lett. {\bf 30}, 848 (2004b)].

6. V. V. Bobylev, G. A. Gontcharov, and A. T. Bajkova, Astron. Zh.
{\bf 83}, 821 (2006).

7. F. Comer\'{o}n, Astron. Astrophys. {\bf 351}, 506 (1999).

8. F. Comer\'{o}n and J. Torra, Astron. Astrophys. {\bf 261}, 94
(1992).

9. F. Comer\'{o}n and J. Torra, Astron. Astrophys. {\bf 281}, 35
(1994).

10. W. Dehnen and J.J. Binney, Mon. Not. R. Astron. Soc. {\bf
298}, 387 (1998).

11. M. Duflot, P. Figon, and N. Meyssonier, Astron. Astrophys.,
Suppl. Ser. {\bf 114}, 269 (1995).

12. Yu. N. Efremov, Sites of Star Formation (Nauka, Moscow, 1989)
[in Russian].

13. Y.N. Efremov, Astron. Astrophys. Trans. {\bf 15}, 3 (1998).

14. Y. N. Efremov and B. G. Elmegreen, Mon. Not. R. Astron. Soc.
{\bf 299}, 588 (1998).

15. B. Famaey, A. Jorissen, X. Luri, et al., Astron. Astrophys.
{\bf 430}, 165 (2005).

16. J. Franko, G. Tenorio-Tagle, P. Bodenheimer, et al.,
Astrophys. J. {\bf 333}, 826 (1988).

17. G. A. Gontcharov, Pis'ma Astron. Zh. {\bf 32}, 759 (2006).

18. B. A. Gould, Proc. Am. Ass. Adv. Sci. {\bf 115} (1874).

19. P. Guillout, M.F. Sterzik, J.H.M.M. Schmitt, et al., Astron.
Astrophys. {\bf 337}, 113 (1998).

20. J. F.W.Herschel, Results of Astronomical Observations Made
During the Years 1834, 5, 6, 7, 8 at the Cape of Good Hope (Smith,
Elder and Co., London, 1847).

21. E. Hog, C. Fabricius, V.V. Makarov, et al., Astron. Astrophys.
{\bf 355}, L27 (2000).

22. P. O. Lindblad, J. Palou\^{s}, K. Loden, et al., HIPPARCOS
Venice'97, Ed. by B. Battrick (ESA Publ. Div., Noordwijk, 1997),
p. 507.

23. P. O. Lindblad, Astron. Astrophys. {\bf 363}, 154 (2000).

24. V. V. Makarov, Astron. J. {\bf 126}, 1996 (2003).

25. E. E. Mamajek, Astrophys. J. {\bf 634}, 1385 (2005).

26. E. E. Mamajek, M. Meyer and J. Liebert, Astron. J. {\bf 124},
1670 (2002).

27. B. Nordstr\"{o}m, M. Mayor, J. Andersen, et al., Astron.
Astrophys. {\bf 418}, 989 (2004).

28. K. F. Ogorodnikov, Dynamics of Stellar Systems (Fizmatgiz,
Moscow, 1965; Pergamon, Oxford, 1965).

29. C. A. Olano, Astron. Astrophys. {\bf 112}, 195 (1982).

30. C. A. Olano, Astron. Astrophys. {\bf 121}, 295 (2001).

31. V. G. Ortega, R. de la Reza, E. Jilinski, et al., Astrophys.
J. {\bf 609}, 243 (2004).

32. C. A. Perrot and I. A. Grenier, Astron. Astrophys. {\bf 404},
519 (2003).

33. A. E. Piskunov, N. V. Kharchenko, S. R\"{o}ser et al., Astron.
Astrophys. {\bf 445}, 545 (2006).

34. S. B. Popov,M. Colpi, M. E. Prokhorov, et al., Astron.
Astrophys. {\bf 406}, 111 (2003).

35. W.G. L. P\"{o}ppel and P.Marronetti, Astron. Astrophys. 358,
299 (2000).

36. N. Robichon, F.A. Arenou, J.-C. Mermilliod, et al., Astron.
Astrophys. {\bf 345}, 471 (1999).

37. M. J. Sartori, J. R. D. Lepine, and W. S. Dias, Astron.
Astrophys. {\bf 404}, 913 (2003).

38. J. Skuljan, J.B. Hearnshaw, and P.L. Cottrell, Mon. Not. R.
Astron. Soc. {\bf 308}, 731 (1999).

39. D. M. Sfeir, R. Lallement, F. Grifo, et al., Astron.
Astrophys. {\bf 346}, 785 (1999).

40. I. Song, B. Zuckermann, and M.S. Bessel, Astrophys. J. {\bf
599}, 342 (2003).

41. J. Torra, D. Fern\'andez, and F. Figueras, Astron. Astrophys.
{\bf 359}, 82 (2000).

42. N.V. Kharchenko, Kinematika Fiz. Nebesnykh Tel {\bf 17}, 409
(2001).

43. N. V. Kharchenko, A. E. Piskunov, S. R\"{o}ser, et al.,
Astron. Astrophys. {\bf 438}, 1163 (2005a).

44. N. V. Kharchenko, A. E. Piskunov, S. R\"{o}ser, et al. Astron.
Astrophys. {\bf 440}, 403 (2005b).

45. The HIPPARCOS and Tycho Catalogues, ESA SP-1200 (1997).

46. R. Wichmann, J.H.M.M. Schmitt, and S. Hubrig, Astron.
Astrophys. {\bf 399}, 983 (2003).

47. P. T. de Zeeuw, R. Hoogerwerf, J. H. J. de Bruijne, et al.,
Astron. J. {\bf 117}, 354 (1999).

}

\bigskip
Translated by V. Astakhov

\begin{table*}[t]                                                
\caption[]{\small\baselineskip=1.0ex\protect
 Space velocities of open clusters and OB associations
corrected for the differential Galactic rotation
 }
\begin{center}
\begin{tabular}{|l|c|r|r|r|c|r|c|c|c|c|c|}\hline
 Cluster
 & $d$,~pc
 & $U$,~km s$^{-1}$~~
 & $V$,~km s$^{-1}$~~
 & $W$,~km s$^{-1}$~~ & $n$ & $V_r$,~km s$^{-1}$~~ \\\hline
 Makarov     & $ 14\pm4$  & $ -9.67\pm2.04$ & $-21.06\pm2.22$ & $ -6.40\pm1.81$ & 23 & $ -2.26\pm2.61$\\
 $\beta$ Pic
             & $ 31\pm3$  & $-13.08\pm1.84$ & $-15.66\pm2.02$ & $ -8.20\pm1.53$ & 6  & $  7.47\pm1.93$\\
 Tuc/Hor     & $ 45\pm3$  & $-10.25\pm0.61$ & $-21.60\pm1.57$ & $ -4.91\pm1.35$ & 8  & $ 19.66\pm1.90$\\
 TWA         & $ 68\pm7$  & $ -7.62\pm1.33$ & $-16.59\pm0.92$ & $ -4.25\pm0.80$ & 22 & $  9.60\pm0.74$\\
 US a        & $126\pm2$  & $  1.36\pm0.49$ & $-15.98\pm1.64$ & $ -4.32\pm0.76$ & 42 & $  1.78\pm0.33$\\
 US b        & $173\pm5$  & $  2.00\pm0.55$ & $-19.07\pm2.05$ & $ -4.53\pm1.02$ & 42 & $  2.45\pm0.29$\\
 UCL a       & $122\pm2$  & $ -3.15\pm1.01$ & $-18.62\pm1.58$ & $ -4.10\pm0.74$ & 66 & $  4.13\pm0.47$\\
 UCL b       & $170\pm3$  & $ -4.21\pm1.20$ & $-22.43\pm1.97$ & $ -5.90\pm1.00$ & 75 & $  4.40\pm0.40$\\
 LCC a       & $101\pm1$  & $ -7.55\pm1.32$ & $-16.68\pm0.83$ & $ -6.39\pm0.72$ & 41 & $  9.06\pm0.36$\\
 LCC b       & $133\pm3$  & $ -8.07\pm1.44$ & $-17.91\pm0.90$ & $ -6.74\pm0.90$ & 58 & $  8.84\pm0.38$\\
 Cham        & $145\pm16$ & $ -6.95\pm2.27$ & $-17.22\pm3.52$ & $ -8.35\pm1.35$ & 9  & $ 12.12\pm4.11$\\
 Lac OB1~a   & $224\pm15$ & $ -1.66\pm1.42$ & $ -8.54\pm3.99$ & $ -1.15\pm1.54$ & 10 & $ -8.52\pm4.20$\\
 Lac OB1~b   & $440\pm20$ & $ -3.76\pm1.51$ & $-12.85\pm1.20$ & $ -3.87\pm1.51$ & 23 & $-12.47\pm1.17$\\
 Per OB2     & $315\pm23$ & $-20.29\pm2.29$ & $ -5.95\pm2.03$ & $ -7.93\pm1.63$ & 19 & $ 15.94\pm2.29$\\
 Tr 10       & $375\pm33$ & $-16.75\pm1.85$ & $-17.50\pm3.90$ & $ -9.56\pm1.32$ & 6  & $ 20.07\pm3.92$\\
 NGC 2516    & $380\pm28$ & $ -9.90\pm1.48$ & $-23.02\pm0.42$ & $ -3.88\pm0.98$ & 6  & $ 22.00\pm0.31$\\
 Col 121     & $465\pm28$ & $-12.20\pm1.81$ & $-11.80\pm2.17$ & $ -7.06\pm1.37$ & 20 & $ 23.22\pm2.43$\\
 Cep OB2     & $501\pm36$ & $  0.02\pm1.95$ & $-14.71\pm2.30$ & $ -2.90\pm1.94$ & 15 & $-17.07\pm2.31$\\
 ASCC 16     & $460$      & $-14.22\pm4.09$ & $ -7.96\pm1.67$ & $ -5.39\pm1.53$ & 3   & $ 20.80\pm4.60$\\
 ASCC 18     & $500$      & $-16.27\pm10.1$ & $ -9.32\pm4.15$ & $ -5.80\pm3.70$ &2$^*$& $ 23.8\pm11.5$ \\
\hline
\end{tabular}
\end{center}
 {\small
 Note. n is the number of stars used to calculate the mean radial
velocity.}
\end{table*}

{
\begin{table*}[t]                                                
\caption[]{\small\baselineskip=1.0ex\protect
Space velocities of
open clusters corrected for the differential Galactic rotation
 }
\begin{center}
\begin{tabular}{|l|c|r|r|r|c|r|c|c|c|c|c|}\hline
 Cluster
 & $d$,~pc
 & $U$,~km s$^{-1}$~~
 & $V$,~km s$^{-1}$~~
 & $W$,~km s$^{-1}$~~ & $n$  \\\hline
 Mamajek 1   & $105$ & $-11.09\pm1.65$ & $-18.59\pm1.14$ & $-10.58\pm0.63$ & 2  \\
 Platais 8   & $150$ & $ -9.07\pm1.23$ & $-18.86\pm3.04$ & $ -3.68\pm0.51$ & 5  \\
 IC 2602     & $160$ & $ -4.20\pm1.49$ & $-23.16\pm2.78$ & $ -0.44\pm0.37$ & 12 \\
 IC 2391     & $176$ & $-23.06\pm2.34$ & $-15.60\pm2.51$ & $ -6.12\pm0.64$ & 18 \\
 NGC 2451~a  & $188$ & $-21.47\pm2.18$ & $-12.78\pm4.74$ & $-13.38\pm1.28$ & 8  \\
 Col 65      & $310$ & $-13.59\pm1.85$ & $-11.10\pm0.97$ & $ -6.85\pm0.60$ & 9  \\
 Col 135     & $319$ & $-10.13\pm1.12$ & $ -6.90\pm2.05$ & $-13.86\pm1.29$ & 4  \\
 NGC 2232    & $325$ & $ -7.74\pm3.71$ & $ -8.84\pm2.58$ & $ -9.68\pm1.10$ & 3  \\
 Platais 6   & $348$ &$-17.28\pm11.69$ & $-12.88\pm5.55$ & $-12.53\pm1.78$ & 2$^*$\\
 ASCC 127    & $350$ & $-14.12\pm2.03$ & $-10.43\pm2.55$ & $ -8.42\pm1.17$ & 2  \\
 ASCC 19     & $350$ & $-12.30\pm3.45$ & $ -8.81\pm1.66$ & $ -6.09\pm1.38$ & 4  \\
 IC 4665     & $352$ & $ -5.98\pm1.59$ & $-14.39\pm1.29$ & $ -7.50\pm0.80$ & 17 \\
Stephenson 1 & $373$ & $-13.29\pm1.96$ & $-19.50\pm3.93$ & $-10.20\pm1.59$ & 2  \\
 Col 70      & $391$ & $-12.05\pm1.90$ & $ -8.89\pm0.96$ & $ -5.81\pm0.71$ & 21 \\
 Alessi 5    & $398$ & $-12.25\pm3.34$ & $-20.59\pm8.38$ & $ -7.17\pm1.24$ & 2$^*$\\
$\sigma$ Ori & $399$ & $-20.49\pm0.91$ & $-16.13\pm1.34$ & $ -3.47\pm1.80$ & 2  \\
 NGC 1976    & $399$ & $-18.35\pm2.30$ & $-16.00\pm1.50$ & $ -7.11\pm1.10$ & 24 \\
 NGC 1981    & $400$ & $-19.85\pm2.06$ & $-11.65\pm1.34$ & $ -6.53\pm1.14$ & 5  \\
 Col 140     & $402$ & $-11.68\pm0.87$ & $-13.30\pm0.85$ & $-14.27\pm1.33$ & 3  \\
 Vel OB2     & $411$ & $-11.80\pm1.16$ & $-12.59\pm3.15$ & $ -3.03\pm0.62$ & 6  \\
 Roslund 5   & $418$ & $-16.33\pm1.39$ & $-19.04\pm2.67$ & $ -6.93\pm0.91$ & 11 \\
 NGC 2451~b  & $430$ & $ -9.58\pm2.50$ & $ -5.89\pm6.75$ & $-15.29\pm2.00$ & 3$^*$\\
vdB-Hagen 23 & $437$ & $-13.62\pm1.19$ & $-10.82\pm0.67$ & $ -5.03\pm0.78$ & 2  \\
 Col 69      & $438$ & $-25.32\pm1.37$ & $-11.40\pm0.94$ & $ -7.88\pm0.96$ & 4  \\
 ASCC 20     & $450$ & $-11.89\pm4.28$ & $ -4.95\pm1.74$ & $ -4.96\pm1.51$ & 5  \\
 NGC 2547    & $457$ & $ -6.98\pm0.96$ & $-11.05\pm1.92$ & $-13.82\pm1.37$ & 9  \\
 ASCC 21     & $500$ & $-13.06\pm1.28$ & $ -7.94\pm0.81$ & $ -5.33\pm0.71$ & 7  \\
 NGC 1977    & $500$ & $-12.71\pm2.28$ & $-17.34\pm2.17$ & $ -5.38\pm1.89$ & 3  \\
 NGC 1980    & $550$ & $-13.54\pm1.92$ & $-13.22\pm1.67$ & $ -7.07\pm1.46$ & 5  \\
 \hline
\end{tabular}
\end{center}
 {\small Note. All of the input data were taken from COCD.}
\end{table*}
}


{
\begin{table*}[t]
\caption[]{\small\baselineskip=1.0ex\protect Principal semi-axes
of the residual velocity ellipsoid $\sigma_{1,2,3}$ and their
directions $l_{1,2,3}$ and $b_{1,2,3}$

}
\begin{center}
\begin{tabular}{|c|c|c|c|c|c|}\hline
   & $\sigma_{1,2,3}$,~km s$^{-1}$ & $l_{1,2,3}$ & $b_{1,2,3}$ \\\hline
 1 & $(7.07,4.30,3.18)\pm(0.56,0.51,0.35)$
   & $153^\circ$, $241^\circ$, $130^\circ$
   & $-14^\circ$, $~~5^\circ$, $~75^\circ$ \\\hline
 2 & $(5.29,4.28,2.75)\pm(0.52,0.36,0.42)$
   & $188^\circ$, $268^\circ$, $120^\circ$
   & $-15^\circ$, $~32^\circ$, $~54^\circ$ \\\hline
\end{tabular}\end{center}

\end{table*}}

\newpage
\begin{figure*}[t]
{\begin{center}
  \includegraphics[width=140mm]{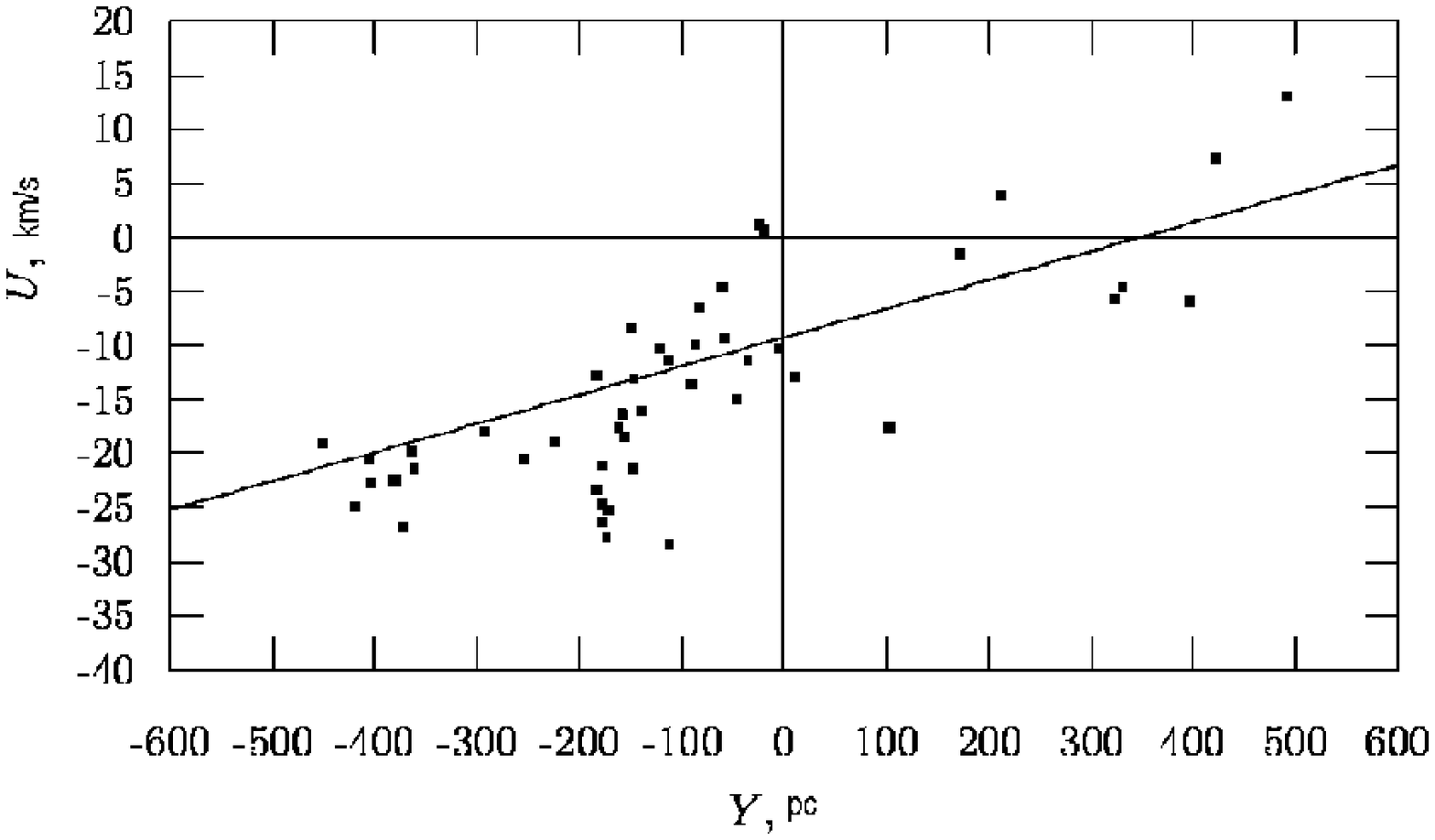}
\end{center}
{{\bf Fig. 1.}
 Dependence of velocity $U$ on coordinate $y$ reflecting
 the influence of the differential Galactic rotation.
 }}
\end{figure*}

\newpage
\begin{figure*}[t]
{\begin{center}
  \includegraphics[width=100mm]{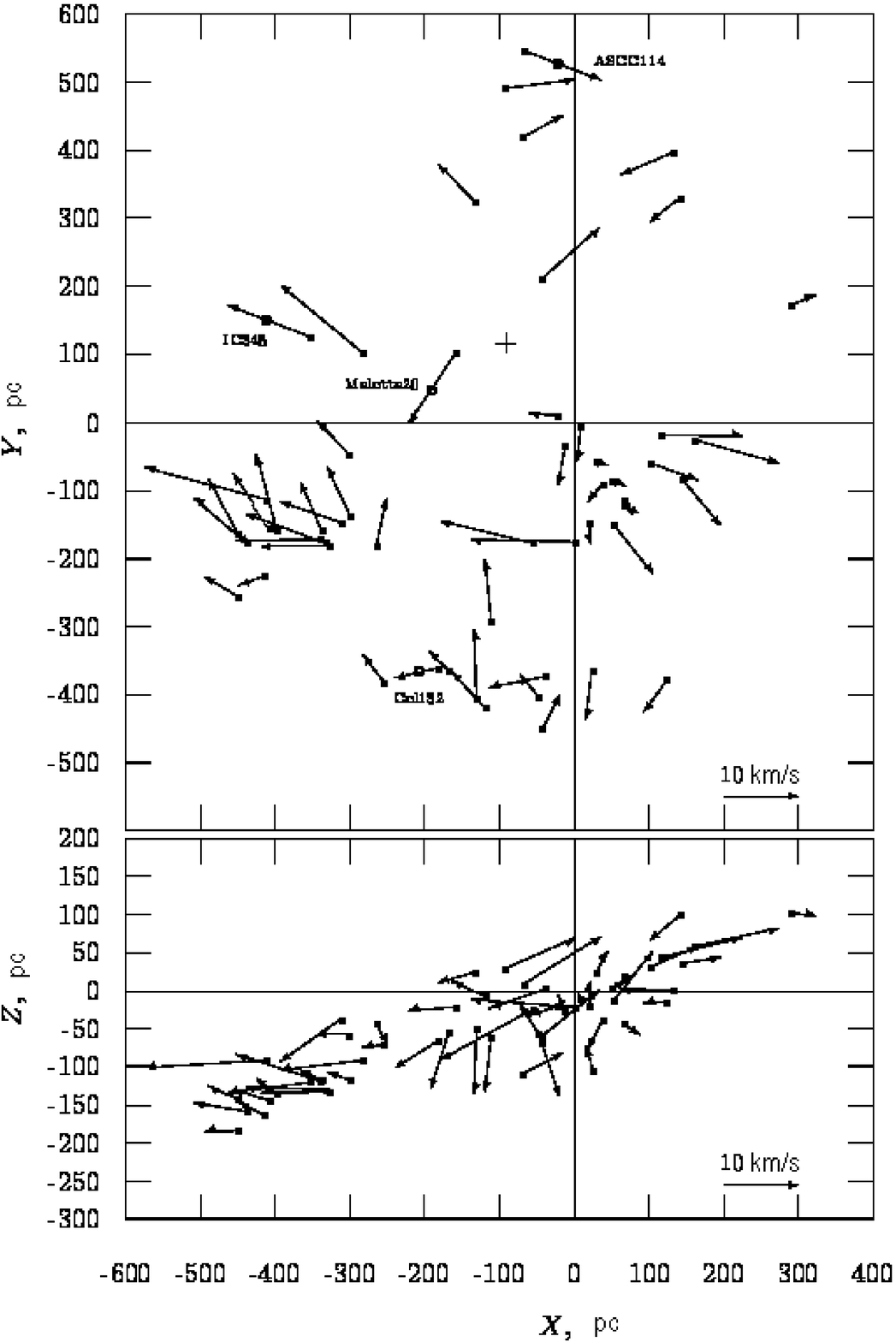}
\end{center}
{{\bf Fig. 2.} (a) Residual velocity vectors $U$ and $V$ in
projection onto the Galactic $xy$ plane; the circles mark the
possible Gould Belt members and the cross marks the kinematic
center $l_\circ=128^\circ$ and $R_\circ=150$ pc. (b) Residual
velocity vectors $U$ and $W$ in projection onto the Galactic $xz$
plane. }}
\end{figure*}

\newpage
\begin{figure*}[t]
{\begin{center}
  \includegraphics[width=140mm]{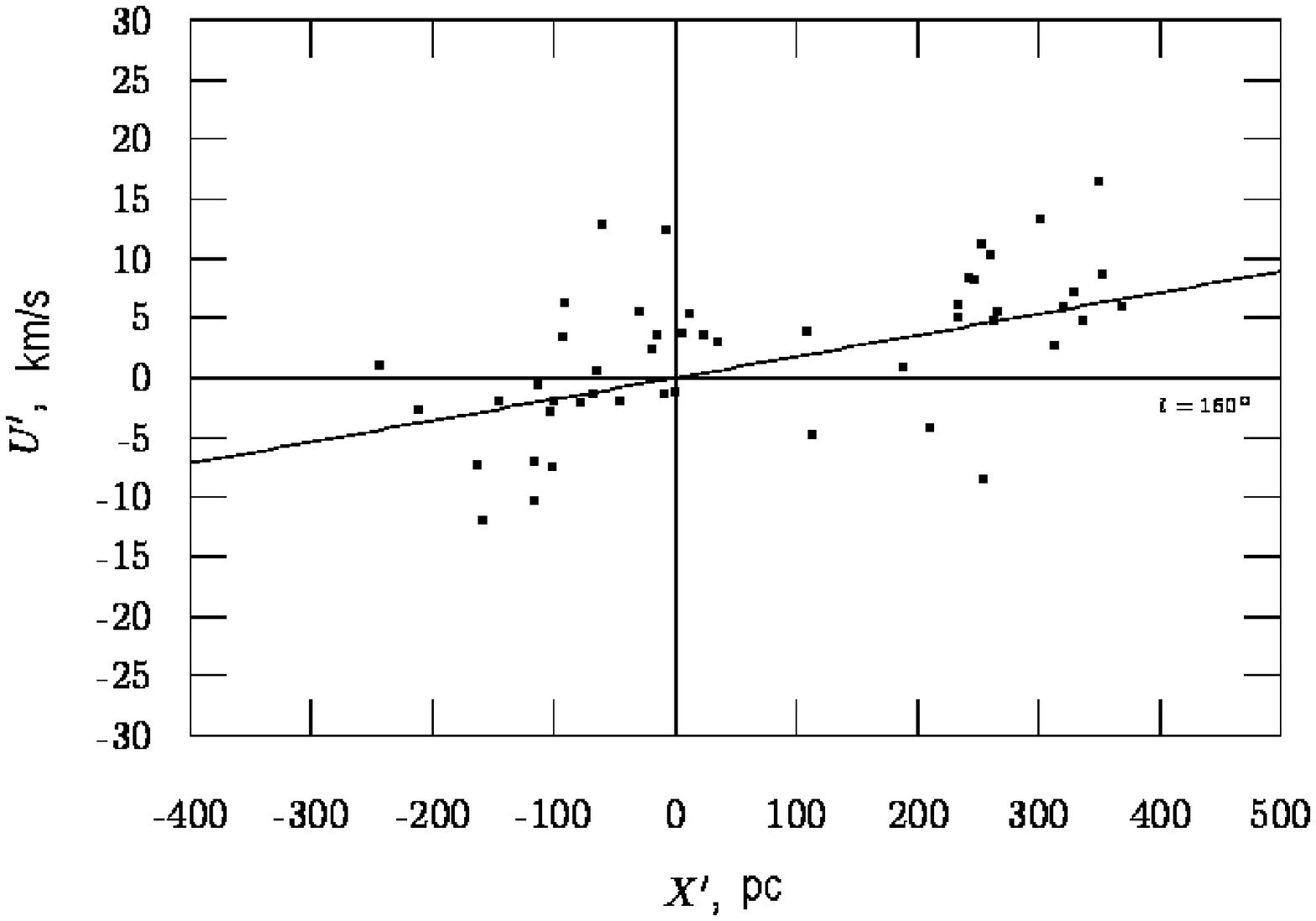}
\end{center}
{{\bf Fig. 3.} Residual velocities $U'$ vs. coordinate $x'$
calculated at $l_\circ=160^\circ$. }}
\end{figure*}

\newpage
\begin{figure*}[t]
{\begin{center}
  \includegraphics[width=70mm]{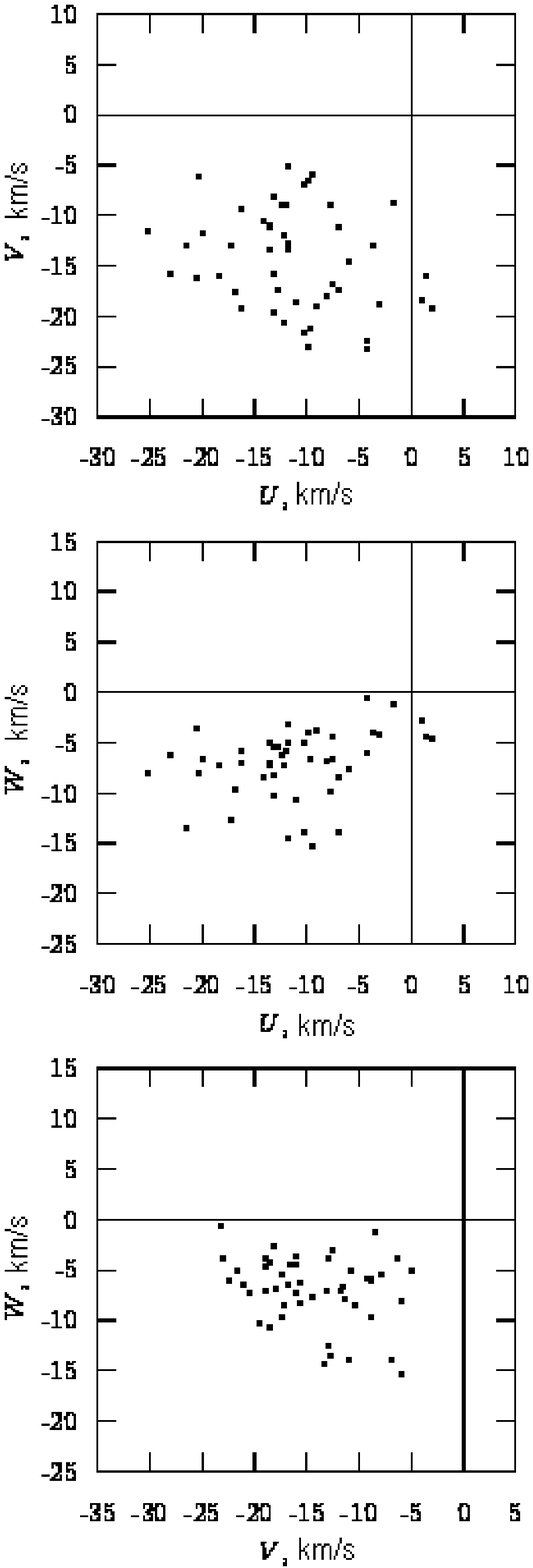}
\end{center}
{{\bf Fig. 4.} Two-dimensional residual velocity $UVW$
distributions for the open cluster complex belonging to the Gould
Belt. }}
\end{figure*}

\newpage
\begin{figure*}[t]
{\begin{center}
  \includegraphics[width=140mm]{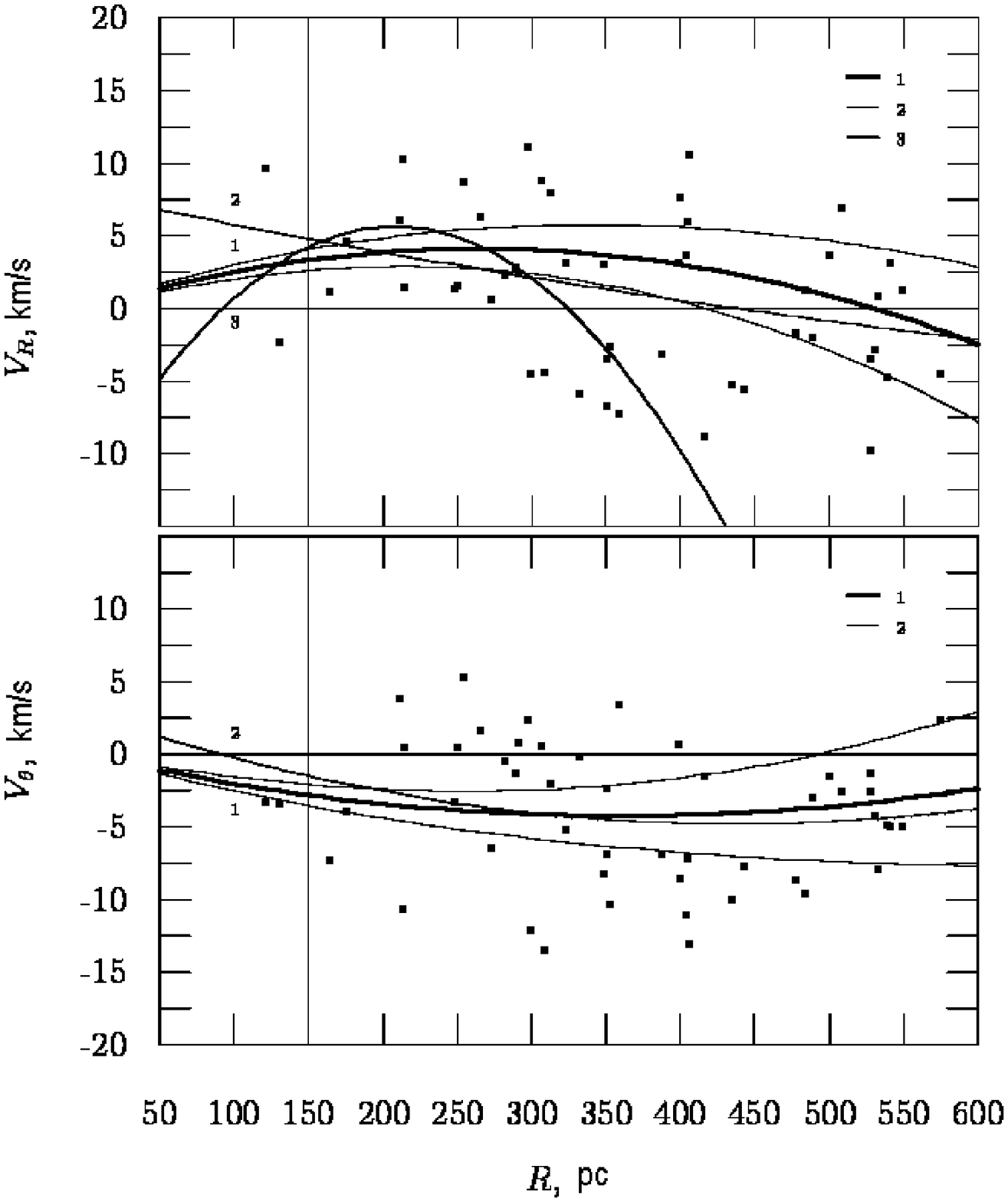}
\end{center}
{{\bf Fig. 5.} Residual velocities $V_R$ (a) and $V_\theta$ (b)
vs. distance from the kinematic center R calculated at
$l_\circ=128^\circ$ and $R_\circ=150$ pc; the vertical line marks
$R_\circ$. See also the text. }}
\end{figure*}

\newpage
\begin{figure*}[t]
{\begin{center}
  \includegraphics[width=140mm]{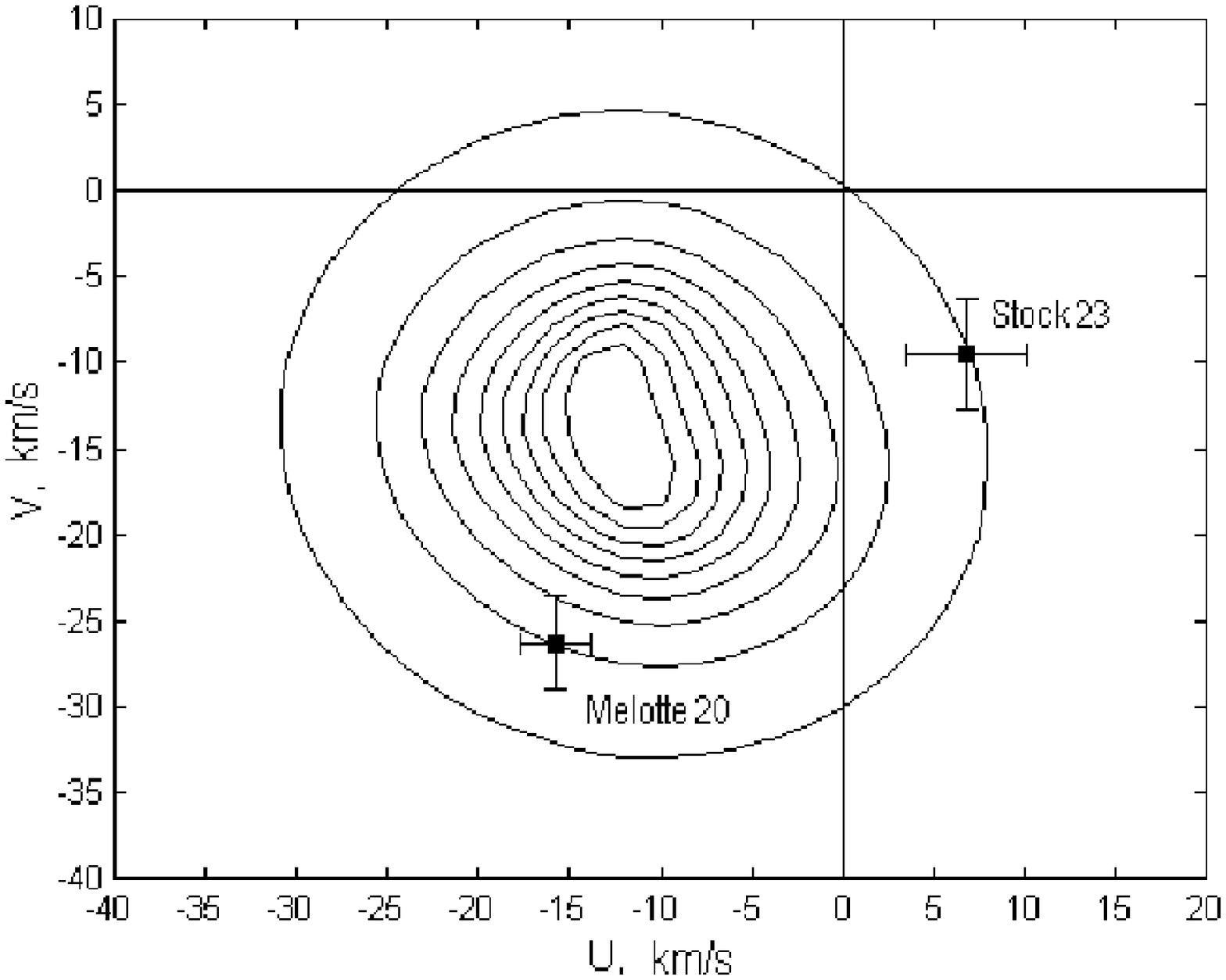}
\end{center}
{{\bf Fig. 6.} Smoothed distribution of residual $UV$ velocities.
}}
\end{figure*}

\end{document}